# UV Light Shower Simulator for Fluorescence and Cerenkov Radiation Studies

P. Gorodetzky[a], J. Dolbeau[a], T. Patzak[a], J. Waisbard[a], C. Boutonnet[a], J.J. Jaeger[a].
*(a) APC-College de France, 11 place Marcelin Berthelot, F-75231 Paris Cedex 05, France\**
Presenter: P. Gorodetzky (philippe.Gorodetzky@cern.ch), fra-gorodetzky-P-abs1-HE1.5-poster

All experiments observing showers light use telescopes equipped with pixellised photodetectors. Monte-Carlo (MC) simulations of the apparatus operation in various situations (background light, shower energy, proximity of tracks...) are mandatory, but never enter into detector details like pulse shape, dead-time, or charge space effects which are finally responsible for the data quality.
An apparatus where each pixel receives light from individual 370 nm UV LEDs through silica fibers is being built. The LEDs receive voltage through DACs, which get their input (which pixel, at what time, which amplitude) from a shower plus noise generator code. The typical time constant of a shower being one µs (300 m for light), the pulses are one µs wide. This is rather long compared to the intrinsic time constant (around 10 ns) of the light detectors, hence, these see "constant light" changing every µs. This is where important loading effects which are not included in MC code can be observed.
The fibers illuminate the pixels through a diffuser, and each fiber illuminates only one pixel.
The number of equipped pixels is such that it englobes a full shower (much less than the full focal surface).
Finally, this equipment can be used also to calibrate the pixels.

## 1. Introduction

One important way to study Cosmic Rays (CR) showers is to look at the fluorescence light emitted by air nitrogen when it is excited by the charged particles of the shower. One can look from the ground (e.g. AUGER, HIRES…) or from space (EUSO). The primary particle can be charged ("ordinary" CR) or neutral: In that case, the shower is mainly electromagnetic. If the energy is above $10^{18}$ eV, there is enough light to see an inclined shower: that is a shower not directed towards the telescope. This is how very high energy neutrinos or gamma-rays are looked for. However, for lower energies, the only hope to see the shower is on its axis through Cerenkov light. In that case, because the shower develops at the same velocity than the light, the observed light pulse is very narrow (a few tens of ns). But aside from that extreme case, most showers do not direct towards the observer. Because the shower length is a few kilometers, the time distribution is of some tens of µs. This is illustrated in Fig. 1. for a shower at 60° seen by EUSO where the full development of the shower is about 150 µs. One can also see that Cerenkov light, not impinging directly on the detector, but Rayleigh and Mie scattered, is an important contribution to the total light. The elementary time length for such an experiment is the µs. This is about 100 times longer than the PMT time constant (10 ns). One can say that the PMT works in DC mode.
The other important effect visible on Fig. 1. is the background due to star light reflected from earth or clouds: 5 photons per µs, or $10^6$ photo electrons (pe) per pixel (1 km$^2$ on ground) per second. This is huge. Furthermore, one PMT has 36 pixels, and works with a gain of $10^6$. The loading on the PMT is far from negligible.
If a shower ($10^{20}$ eV) is present, some $10^7$ pe per pixel per second add to the background. If one wants the system not to saturate, it has to be able to sustain 10 times that, that is $10^8$ pe per pixel per second. The current flowing in the PMT will be of the order of 30 µA (only a few pixels per PMT see the shower).

Hence, the voltage divider should be able to deliver some 3 mA, or 3 W per PMT. If EUSO has 5000 PMTs, total power would be 15 kW. This is impossible. The voltage divider has to be very clever, for instance, very high efficiency power supplies for each dynode. However, the obvious problem is to test such extreme conditions, which is the reason why this simulator is built.

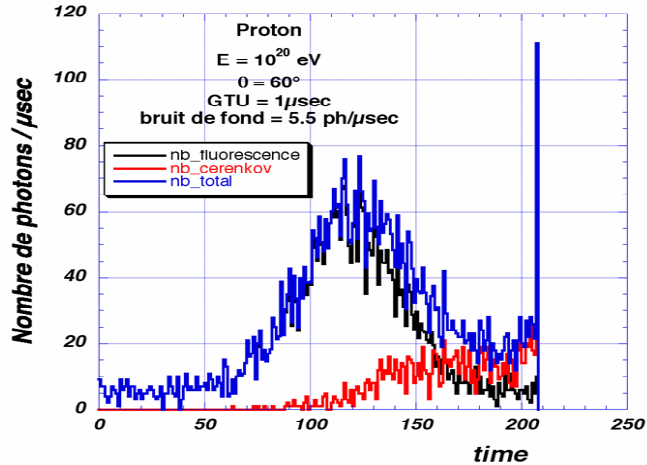

**Figure 1.** Left: time distribution of the photons per pixel arriving at the EUSO focal surface for a proton shower of E=$10^{20}$ eV with an incidence angle of 60°. The fluorescence and Cerenkov contributions are distinguished. The black histogram corresponds to the photons flux if the atmosphere was totally transparent. A GTU is a Gate Time Unit.

## 2. The UV Simulator.

The basic problem is to feed DC light (changing slowly at the rate of a few per cent per µs) to the pixels in a controlled way. The pixels have to be independent. Let us illustrate the case of a EUSO 36 pixel (6x6) R8900-03-M36 Hamamatsu PMT. Thirty six UV (370 nm) LEDs send light through quartz fibers to the PMT. Light exiting the fibers go to the pixels through a 0.5 mm thick Teflon® sheet to uniformize the light evenly on the pixel surface. Fig. 2. shows the mechanical arrangement for one pixel of the PMT. An advantage of the use of fibers is that the LEDs do not have to be in the same arrangement as the pixels, but are conveniently set on a printed circuit board. Figure 3. shows the LED characteristics when DC voltage is varied, and when the voltage passes through a linear gate to be set to 0 for one µs every 2 µs (see the legend of figure 2).
With the basic voltage divider shown in fig.2, and sending light on one pixel only, the PMT saturates already at a rate of $5 \cdot 10^6$ pe / sec which is much lower than required. Just the background noise on the 36 pixels would be around $3 \cdot 10^7$ / sec!

A Monte-Carlo program produces a number of shower photons for every µs of its development, analogous to the Fig 1 (it naturally includes the noise). These numbers are converted into volts using the LED characteristics of Fig 3. The resulting table of volts is sent through a USB interface on Digital to Analogous Converter (DAC) which will yield the "real" voltages applied to the LEDs. Their rise and fall time are brief: less than 100 ns. The voltage out of the DAC goes through a linear gate before going to the LED. The computer triggers the DACs and the gates in real shower times. Every µs a new order is given to the linear gate.

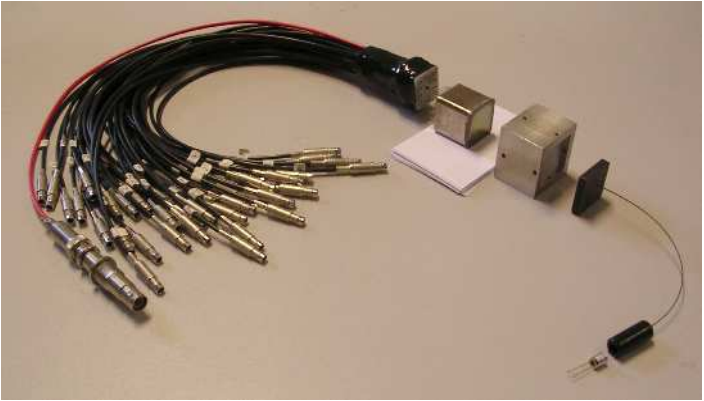 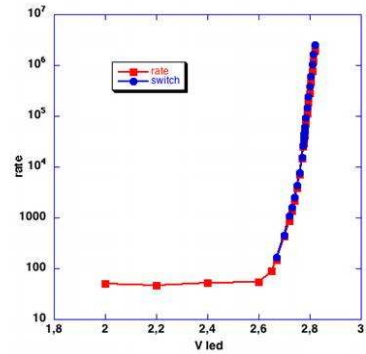

**Figures 2.**& **3**. Exploded view of the arrangement. Down right, one can see the UV LED, followed by the coupler to the fiber. At the other extremity, the fiber goes into a 2-D comb. The intermediate aluminum piece couples exactly the fibers to the PMT which is seen next. This coupler has square holes with very thin walls, corresponding precisely to the PMT pixels. The Teflon sheet goes between the black comb and the coupler. It is pressed by the black comb against the square holes of the coupler. Finally, the Hamamatsu voltage divider is shown. The **figure 3** on the right shows the photoelectrons rate in photoelectrons per second when the voltage applied to the LED is varied: red curve shows a DC voltage applied, and single pe are counted. In the blue curve the DC voltage is goes through a linear gate where it is set to 0 for 1 μs every 2 μs. The number of pe should then be exactly half of the number obtained without the gate. In the figure, the number of pe corresponding to the gate has been multiplied by two to take care of that effect. Hence, the blue curve is exactly on the red one, illustrating the system linearity.

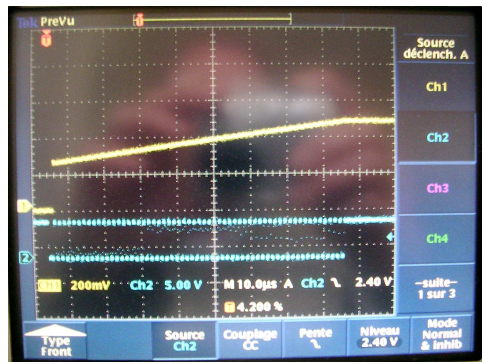

**Figure 4.** Voltages (white track) created with a DAC controlled by the computer. Here, a ramp followed by a plateau was made.

Fig. 4 shows an oscilloscope readout of a DAC to where the computer asked to make a ramp followed by a plateau.

## 2. Discussion

Shown in Fig. 4 the way a shower should be seen in a computer simulation with the time in ordinate and X (Y) of the pixels of the focal surface in abscissa. Noise has been added to the shower. The aim is to see the same figure with the PMT outputs and applying the correct triggers.

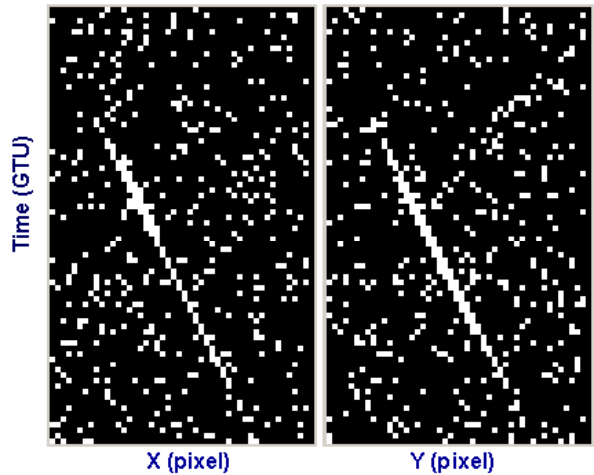

**Figure 5.** Schematic view of a track along the two projection view (as registered by the read-out electronics). Shower simulated at energy $E = 10^{20}$ eV and arrival direction zenith angle ~ 60°.

## 3. Conclusions

The intense background light due to the stars induces a tremendous load on the PMTs used for CR measurement through fluorescence detection. This light is DC, and even the shower light itself can also be considered as DC. This apparatus allows to check that the PMTs are used in a mode where they are still linear.

## 4. Acknowledgements

We thank the EUSO Japanese group at Riken for lending us the Hamamatsu PMT, E. Plagnol (APC, Paris) who provided fig. 1 and O. Catalano (IASF, Palermo) who made the simulation shown fig. 5.